\documentstyle[amsfonts,epsfig,portland]{aipproc}
\begin{document}
\title{From known to undiscovered resonances}
\author{Bijan Saghai}
\address{Service de Physique Nucl\' eaire, DAPNIA, CEA/Saclay, 
91191 Gif-sur-Yvette, France, email:~bsaghai@cea.fr}
\maketitle
\begin{abstract}
Electromagnetic meson production formalisms are reviewed, with emphasise 
placed on their ability in search for new baryon resonances {\it via}
$\gamma p \rightarrow K^{+} \Lambda$ and
$\gamma p \rightarrow \eta p$ processes. 
The relevant studies, aiming to deepen our insights to hadron spectroscopy, 
constitute strong tests of the QCD inspired theoretical developements.
\end{abstract}
\section{Introduction}
presently, our knowledge on the baryon resonances comes~\cite{PDG,PWA} 
mainly from 
partial wave analysis of the ``pionic'' 
processes $\pi N \to \pi N ,~\eta N$,  $\gamma N \to \pi N$, and to less
extent, from two pion final states.

The advent of new facilities 
offering high quality electron and photon beams and 
sophisticated detectors, has stimulated intensive experimental
and theoretical study of the mesons photo- and electro-production.
One of the exciting topics here is the search for new baryon resonances 
which do not couple or couple too weakly to the $\pi N$ channel. 
%
Several such resonances have been  predicted~\cite{Miss-res} by different QCD 
inspired approaches, offering strong test of the underlying concepts.

In this note, we concentrate on the interpretation of pseudoscalar mesons 
photoproduction recent data~\cite{Saphir,Mainz,elsa,graal,Graal2000}, 
where manifestations of new resonances were reported~\cite{MB,LS-3}. 
The processes under consideration, $\gamma p \rightarrow K^{+} \Lambda$ and
$\gamma p \rightarrow \eta p$, are basically studied {\it via} two families of
formalisms:
\begin{itemize}
\vspace{-2.mm}
\item{
Effective Lagrangian approach, where the amplitudes are in
general expressed as Feynman diagrams at tree level.
}
\vspace{-2.mm}
\item{
Constituent quark approach based on the broken $SU(6)\otimes O(3)$ 
symmetry.
}
\vspace{-2.mm}
\end{itemize}

Below, we summarize the basis of these formalisms and
examin their findings with respect to the reported new resonances.
%
%
\section{Theoretical Frames}
%
%
For several decades, effective Lagrangian family approaches have been extensively
developped and applied to the electromagnetic production of pseudoscalar mesons.
%
%
\begin{table}[t!]
\begin{center}
\begin{tabular}{cll}
\hline 
 Baryon &  Three and four star resonances &  One and two star resonances \\
\hline
 $N^*$ & $S_{11}(1535)$, $S_{11}(1650)$, & $S_{11}(2090)$, \\
 & $P_{11}(1440)$, $P_{11}(1710)$, $P_{13}(1720)$,
 & $P_{11}(2100)$, $P_{13}(1900)$, \\
 & $D_{13}(1520)$, $D_{13}(1700)$, $D_{15}(1675)$, 
 & $D_{13}(2080)$, $D_{15}(2200)$, \\
 & $F_{15}(1680)$,  
 & $F_{15}(2000)$, $F_{17}(1990)$, \\
 & $G_{17}(2190)$, $G_{19}(2250)$,  
 &   \\
 & $H_{19}(2220)$, 
 &   \\ [10pt]
 $\Lambda^*$ 
 & $S_{01}(1405)$, $S_{01}(1670)$, $S_{01}(1800)$,  
 &  \\
 &  $P_{01}(1600)$, $P_{01}(1810)$, $P_{03}(1890)$, 
 & \\
 & $D_{03}(1520)$, $D_{03}(1690)$, $D_{05}(1830)$,
 & $D_{03}(2325)$, \\
 & $F_{05}(1820)$, $F_{05}(2110)$,  
 & $F_{07}(2020)$,  \\
 & $G_{07}(2100)$, 
 &   \\
 & $H_{09}(2350)$,
 &   \\ [10pt]
 $\Sigma^*$ 
 & $S_{11}(1750)$, 
 & $S_{11}(1620)$, $S_{11}(2000)$, \\
 & $P_{11}(1660)$, $P_{11}(1880)$, $P_{13}(1385)$, 
 & $P_{11}(1770)$, $P_{11}(1880)$, $P_{13}(1840)$, \\ 
 && $P_{13}(2080)$, \\ 
%
 & $D_{13}(1670)$, $D_{13}(1940)$, $D_{15}(1775)$,
 & $D_{13}(1580)$, \\
 & $F_{15}(1915)$, $F_{17}(2030)$. 
 & $F_{15}(2070)$,   \\
 &  
 & $G_{17}(2100)$.  \\
\hline
\end{tabular}
\caption{Isospin-1/2 baryon resonances~[1] with mass 
$M_{N^*}\leq$ 2.5 GeV. Notations are $L_{2I~2J}(mass)$ and 
$L_{I~2J}(mass)$ for $N^*$ and $Y^*$, respectively.}
\end{center}
\vspace{-8mm}
\end{table}
%
%
%

The most studied channel is by far, the single pion photoproduction where, in the
investigated kinematic regions, the reaction mechanism is dominated by the $\Delta$(1234)
resonance. Such a feature has also been observed, although not fully understood, in the
case of the $\eta$ meson production, where the $S_{11}$(1535) resonance 
plays a dominant role,
at least up to $\approx$ 100 MeV above threshold. However, the associated strangeness
production channel has not shown any strong preference for a given resonance.

The recent data from high duty cycle accelerators allow a real break through in this
field and extend the measured (measurable) domains well above threshold and give
acces to polarization observables. Then, the relevant formalisms need to have the
ability of incorporating a large number of resonances summarized in Table~1.
This requirement becomes crucial in searching for new baryon resonances, 
on which this note focuses.

In this Section, we concentrate on two of the most commonly used formalisms,
namely, tree level diagrammatic effective 
Lagrangians~\cite{MB,TRR,Cohen,NC,Ohio,AS,SL,OS,Yonsei,RPI_1,RPI_2}, 
and constituent quark
approaches~\cite{LS-3,zpl97,LS-1,LST}. For other relevant approaches, the
reader is refered to Refs.~\cite{LS-3,HYP97}.
\vspace{-1mm}
\subsection{Meso-baryonic effective Lagrangian approach}
In lines with single photoproduction formalisms, the effective Lagrangian
approaches have been extended to the 
$KY$~\cite{MB,TRR,Cohen,NC,Ohio,AS,SL,OS,Yonsei} and the $\eta N$~\cite{RPI_2} final states.
In this Section, we limit ourseleves to the former channel.

The history of strangeness physics studies {\it via} electromagnetic probes can
be divided into two periods (see, e.g., Refs.~\cite{AS,HYP97}). 
The early works started in the late 50's and
went on for about 15 years.
Then, in the early 80's, several experimental projects restored this dormant 
field and gave it a promising future, and due to several foreseen facilities with high 
quality polarized electron and/or photon beams, revived 
theoretical investigations in this realm.
The starting point of these studies is the effective hadronic
Lagrangian approach, using diagrammatic techniques, developed 
in the old days by Thom, Renard and Renard~\cite{TRR}.  
However, these works
led to a confusing situation~\cite{AS} on the ingredients of the 
elementary operator.
These pioneering attempts were followed by more extensive 
investigations~\cite{Cohen,NC,Ohio,AS,SL,OS,Yonsei} 
of the elementary reactions: 
${\gamma}p~{\to}~K^+ \Lambda,~K^+ \Sigma^{\circ},~K^{\circ} \Sigma^+$,
with real and virtual photons, as well as the crossing symmetry 
channels~\cite{NC,SL,OS} 
$K^- p ~\to~ \gamma  \Lambda,~\gamma  \Sigma^{\circ}$.

In this note, we wish to comment
on the capability of different formalisms in handling the exchanged
resonances in the elementary reaction mechanism.

The most widely used effective
Lagrangians are based on the tree approximation, allowing the
inclusion of a large number of possible exchanged particles
in the {\it s-, u-}, and {\it t-}channels {\it via} the relevant 
Feynman diagrams.
Within such phenomenological approaches, {\it a priori} more than 30 exchanged
baryonic resonances~(Table~1) can intervene. This uncomfortable
situation, where no dominante resonances could be identified, is due to
our lack of knowledge~\cite{PDG} on the photo-excitation couplings and/or
on the branching ratios of these resonances to the relevant KY final states.

This raises a {\it crucial question}: does a given formalism allow us to 
introduce in a model, baryon resonances with spin 1/2, 3/2, and 5/2?
Capabilities of the most commonly used formalisms which deal with the above
question are summarized in Table~2. 
\begin{table}[b!]
\begin{center}
\begin{tabular}{lcccccc}
\hline  
{Resonance~(spin) $\rightarrow$} & $N^*$(1/2) & $N^*$(3/2) & $N^*$(5/2) & $Y^*$(1/2) & $Y^*$(3/2) & \\
{Group [Ref.] $\downarrow$} & & & &  & & Off-shell treatment\\
\hline 
North Carolina [13]& Y &  &  & Y& &  \\
Ohio - GWU [9,14]& Y & Y &  & Y &  & \\
Saclay - Lyon [16]& Y & Y & Y & Y &  & \\
VPI - Lyon - Saclay [17] & Y & Y &  & Y & Y & Y \\
Yonsei [18] & Y & Y & Y & Y & Y &  \\
\hline
\end{tabular}
\caption{ Capabilities of the most commonly used formalisms for the
process $\gamma p \to K^+ \Lambda$ in handling
nucleonic resonances with spin 1/2, 3/2, and 5/2, and hyperonic resonances
with spin 1/2 and 3/2. 
All the models issued from these formalisms include, besides the extended
Born terms, the $K^*$(892) and $K1$(1270) exchanges in the
{\it t-}channel.} 
\end{center}
\end{table}

One of the main  sources of the level of
success of the models built within these formalisms, is the inclusion of the
{\it t-}channel resonances. Actually, we know~\cite{ST} from the 
duality hypothesis
that there is a close relationship between a dynamical model's content with 
respect to the included baryon resonances and the corresponding strength of the 
{\it t-}channel exchanges needed to fit the data. 
However, this artifact does not provide fine enough insights into the reaction 
mechanism: in a given model, we cannot say which baryon resonances, absent in
the model, are mimiked by the {\it t-}channel contributions. However, within
the formalisms discussed in this Section, serious technical difficulties
prevent us from introducing all the known baryon resonances and hence to
discard {\it t-}channel contributions.

A major problems in the formalisms handling baryonic resonances with
spin higher than 1/2 is related to the adopted propagators. 
As shown by the RPI group~\cite{RPI_1}, the most commonly 
used propagator~\cite{Ohio,SL} for spin 3/2 nucleonic resonances
has no inverse. Moreover, this undesirable situation prevents those formalisms
from introducing spin 3/2 hyperonic resonances, which otherwise would lead 
to an unwanted singularity in the $u$-channel.
To overcome this serious shortcoming, 
the RPI group, investigating the pion and $\eta$ photoproduction reactions, has
applied~\cite{RPI_1} the Rarita-Schwinger approach~\cite{RS}. Along the same lines, the authors of
Ref.~\cite{OS} have produced a formalism allowing a proper treatment of {\it both}
nucleonic and hyperonic spin 3/2 resonances. For recent discussions on various 
aspects of this topic see, e.g. Refs~\cite{OS,RPI_1,Pasc,Hemm}.

In Sec. III.A, the results of the formalism discussed above, will be compared
to the $\gamma p \to K^+ \Lambda$ data.
%
%
\subsection{Constituent quark effective Lagrangian approach}
The starting point of the meson electromagnetic production in the chiral quark 
model is the low 
energy QCD Lagrangian~\cite{QCD_L}.
The baryon resonances in the {\it s-} and {\it u-}channels are treated as three
quark systems.
The transition matrix elements based on the low energy QCD
Lagrangian include the {\it s-}, {\it u-}, and {\it t-}channel contributions
${\cal M}_{if}={\cal M}_s+{\cal M}_{u}+{\cal M}_{t}$.
The contributions from  the {\it s-}channel resonances can be written as
\begin{eqnarray}\label{eq:MR}
{\mathcal M}_{N^*}=\frac {2M_{N^*}}{s-M_{N^*}(M_{N^*}-i\Gamma(q))}
e^{-\frac {{k}^2+{q}^2}{6\alpha^2_{ho}}}{\mathcal A}_{N^*},
\end{eqnarray}
where  $k$ and $q$ represent the momenta of the incoming photon 
and the outgoing meson respectively, $\sqrt {s}\equiv W$ is the total c.m. energy of 
the system, $e^{- {({k}^2+{q}^2)}/{6\alpha^2_{ho}}}$ is a form factor 
in the harmonic oscillator basis with the parameter $\alpha^2_{ho}$ 
related to the harmonic oscillator strength in the wave-function, 
and $M_{N^*}$ and $\Gamma(q)$ are the mass and the total width of 
the resonance, respectively.  The amplitudes ${\mathcal A}_{N^*}$ 
are split into two parts~\cite{zpl97}: the contribution 
from each resonance below 2 GeV, the transition amplitudes of which 
have been translated into the standard CGLN amplitudes in the harmonic 
oscillator basis, and the contributions from the resonances above 2 GeV
treated as degenerate, since little experimental information is available
on those resonances.

The {\it u-}channel contributions are divided into the nucleon Born
term and the contributions from the excited resonances.  The matrix 
elements for the nucleon Born term is derived explicitly, while the 
contributions from the excited resonances above 2 GeV for a given parity 
are assumed to be degenerate so that their contributions could be 
written 
in a compact form.

The {\it t-}channel contribution contains two parts: 
{\it i)} charged meson exchanges which are proportional to the charge of outgoing 
mesons and thus do not contribute to the process $\gamma N\to \eta N$;  
{\it ii)} $\rho$- and $\omega$-exchange in the $\eta$ production which are 
excluded here due to the duality hypotheses; as discussed in Ref.~\cite{LS-3}.

Within
the exact $SU(6)\otimes O(3)$ symmetry the $S_{11}(1650)$ and 
$D_{13}(1700)$ do not contribute to the investigated 
reaction mechanism. However, the breaking of this symmetry leads to
the configuration mixings.
Here, the most relevant configuration mixings are~\cite{LS-3} those of the
two $S_{11}$ and the two $D_{13}$ states around 1.5 to 1.7 GeV. The 
configuration mixings, generated by the gluon exchange interactions in 
the quark 
model~\cite{IKK}, can be expressed in terms of the mixing angles,
$\Theta_S$ and $\Theta_D$,
between the two $SU(6)\otimes O(3)$ states $|N(^2P_M)>$  and 
$|N(^4P_M)>$, with the total quark spin 1/2 and 3/2. Results of this
approach will be compared to the data for $K^+ \Lambda$ and $\eta p$
channels in Sections III.A and III.B, respectively.
%
%
\section{Evidence for new resonances?}
\subsection{Associated strangeness production}
Recent SAPHIR data~\cite{Saphir} for the $\gamma p \to K^+ \Lambda$ process 
has been claimed~\cite{MB} to provide {\it evidence} for a missing $D_{13}$ 
resonance~\cite{Capstick}. In their work based on a meso-baryonic effective 
Lagrangian approach (see Sec. II.A), Mart and Bennhold (MB) produce a 
model which contains contributions from;
\vspace{-1mm}
\begin{itemize}
\item{Extended Born terms.}
\vspace{-2mm}
\item{Two kaonic resonances in the {\it t-}channel: $K^*$(892) and $K1$(1270).}
\vspace{-2mm}
\item{Three established nucleonic resonances in the {\it s-}channel:
$S_{11}(1650)$, $P_{11}(1710)$, and $P_{13}(1720)$; hereafter
referred to as $N4 $, $N6 $, and $N7 $, respectively.
Notice that (Table~1) the two first resonances have spin-1/2, while the third one
is a {\it spin-3/2} resonance. The propagators are in line with those
in Ref.~\cite{Ohio} and {\it do not} embody off-shell treatments.}
\vspace{-2mm}
\item{One unknown (or missing) {\it spin-3/2} nucleonic resonance, that the
authors determine as $D_{13}$(1895).}
\vspace{-1mm}
\end{itemize}
It is to be noted that this model has no hyperonic resonances.

Figure~1 shows the total cross section data and the MB complete model
(dash-dotted curve). MB report~\cite{MB} also results with
a model containing all the above ingredients, except the missing resonance.
The main feature of this latter model is that it does not produce any structure around
W=1.9 GeV, as required by the (fitted) data.
%
\begin{figure}[b!]
\begin{center}
\vspace{-8mm}  
\mbox{\epsfig{file=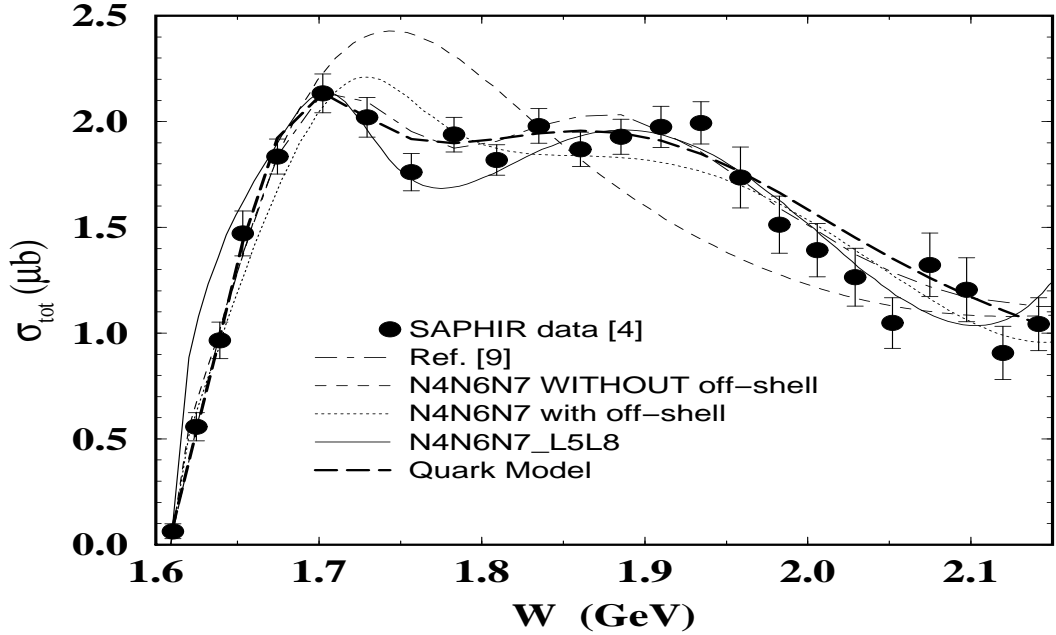,height=9.4cm,width=16.0cm}}
\end{center}
\vspace{-3mm}
\caption{Total cross section for the process $\gamma p \to K^+ \Lambda$ as a function of total center-of-mass energy.
}
\end{figure}
Using the Saclay-Lyon model~\cite{SL}, we~\cite{HYP97} have fitted the same data base
(including differential and total cross-sections from SAPHIR, old recoil
$\Lambda$ asymmetry~\cite{pol}, and recent JLab electroproduction data~\cite{Nicu_98}) 
as MB.
Limiting the reaction mechanism to the above ingredients without 
the missing $D_{13}$ resonance, we obtain the same features as MB. Our fit
for the [N4, N6, N7] set is shown in Fig.~1 (dashed curve) and decreases
monotonically beyond a maximum around W=1.72 GeV. As a next step, and for
the reason explained in Section II.A, we include the off-shell
effect treatments in line with Ref~\cite{OS}. We then get the dotted
curve which shows a significant enhancement in the cross section above 1.85 GeV.
By introducing two hyperonic resonances $P_{01}(1810)$ and 
$P_{03}(1890)$ (hereafter called $L5 $ and $L8 $, respectively), the data
are well reproduced (full curve). This set of resonances [N4, N6, N7, L5, L8]
reproduces reasonably also the other fitted data. 
Our results therefore show that there is {\it no need} for a missing resonance.

The above considerations can not be taken as an attempt to produce a
new model, but just as an {\it illustration} as how cautious we have to be
in using the existing formalisms when searching for new resonances.
A more reliable approach in this respect, should allow us to embody
all known resonances.

Such an opportunity is offered to us by the constituent quark formalims
presented in Section II.B. We~\cite{LST} have included all known nucleon and hyperon
resonances given in Table~1, and have fitted the photoproduction data base.
The result is given in Fig.~1 (heavy dashed curve). 
The agreement between this latter curve
and the data endorses our conclusions that the SAPHIR data does not 
show any manifestations of a new resonance. 
%
\subsection{$\eta$-meson production}
%
Using a constituent quark model (Section II.B), we~\cite{LS-3} have fitted the 
following sets of the $\eta$-photoproduction data: 
differential cross-sections from
MAMI/Mainz~\cite{Mainz} and Graal~\cite{Graal2000}, as well as the polarized 
beam asymmetry from 
Graal~\cite{graal}. Then we have predicted~\cite{LS-3} the total cross-section 
and the polarized target asymmetry. This latter observable has been measured
at ELSA~\cite{elsa}.
%
\begin{figure}[b!]
\begin{center}
\mbox{\epsfig{file=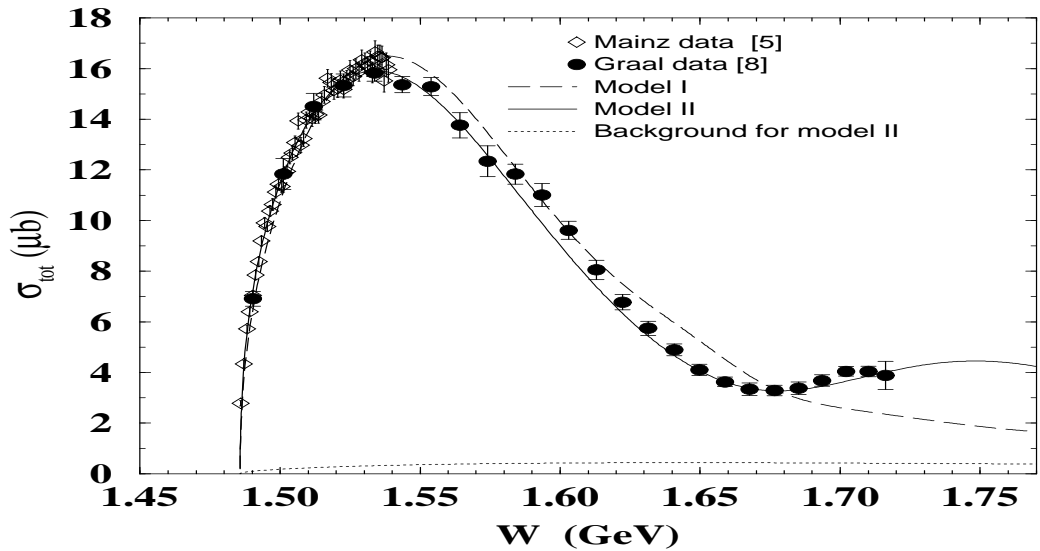,height=8.1cm,width=16.0cm}}
\end{center}
\caption{Total cross section for the reaction 
$\gamma p \to \eta p$
as a function of total center-of-mass energy. 
The curves come from the models I (dashed), II (full). 
The dotted curve shows
the background terms contribution in the model II.
} 
\end{figure}

In Fig.~2, we show comparison for the total cross-section,
for the models I and II.
Both models include all 3 and 4 star known nucleon resonances (Table~1). 
They also satisfy the configuration mixing requirements and for
both models, the extracted mixing angles are in agreement with
the Isgur-Karl~\cite{IKK} predictions. 

The model~I reproduces fairly well the total
cross-section data up to $W \approx$ 1.61 GeV. Between this latter energy and 
$\approx$ 1.68~GeV, the model overestimates the data, and above 1.68~GeV, the
predictions underestimate the experimental results, missing the total cross-section
increase. 

In summary, results of the model I show clearly that an approach containing a
correct treatment of the Born terms and including {\it all known resonances} in the 
{\it s-} and {\it u-}channels {\it does not} lead to an acceptable model, even within
broken $SU(6)\otimes O(3)$ symmetry scheme.
To go further, one possible scenario is to investigate manifestations of yet 
undiscovered resonances.
As already mentioned, rather large number of such resonances has been predicted by 
several authors.
To find out which ones could be considered as relevant candidates, we examined
the available data.

The differential cross-sections~\cite{LS-3,BS_Sendai}, show clearly that this 
mismatch is due to the
forward angle peaking of the differential cross-section
for $W \ge$1.68~GeV ($E_{\gamma}^{lab} \ge$ 1.~GeV). Such a behaviour might
likely arise from missing strength in the $S$-waves. This latter conclusion is
endorsed by the role played by the $E_0^+$ in the multipole structure of 
the differential
cross-section and the single polarization observables.
If there is indeed an additional $S$-wave resonance in this
mass region, its dependence on incoming photon and outgoing meson momenta would be 
qualitatively similar to that of the $S_{11}(1535)$, even though the form factor
might be very different. Thus, for this new resonance, we use the same CGLN
amplitude expressions as for the $S_{11}(1535)$.
We have hence introduced~[1] a third $S_{11}$ resonance and refitted the same data base
as for the model I, leaving it's mass and width as free parameters. 
The results of this model, depicted in Fig.~1 (full curve), reproduce nicely the
data. This is also the case~\cite{LS-3} for the polarized beam and polarized
target asymmetries. For this latter observable, our predictions yield a good
agreement with the data.

For the new $S_{11}$ resonance, we find
M=1.729 GeV and $\Gamma$=183 MeV. These values are amazingly close to those of
a predicted~\cite{zr96} 
third $S_{11}$ resonance, with M=1.712 GeV and $\Gamma_{T}$=184 MeV.
%
%
\section{Summary and concluding remarks}
%
%
In this note we concentrated on the search for new resonances {\it via} the two
processes $\gamma p \to K^+ \Lambda$, $\eta p$ for which recent data have become
available.

The effective Lagrangian approaches, using Feynman diagrams at tree level, 
applied to the above channels, allow to study
some specific aspects of the reaction mechanism. However, they are not suitable
in looking for new resonances. The reason for this shortcoming is that they do not
allow the inclusion of {\it all} relevant known resonances. Although the introduction of spin-1/2
resonances is straightforward, higher spin resonances are more complicated to be
handlded. The main difficulty comes from the
incorporation of the so called {\it off-shell effects}, 
inherent to the fermions with spin $\ge 3/2$. Presently, these effects
can be embodied for spin-3/2 resonances, but no conclusive attempt has been made
for higher spins. 
Another limitation of these approaches is due to the number
of free parameters: one for each spin-1/2, two for each higher spin, plus 3 
off-shell parameters per resonance. In other words, even if we were able to
treat all higher spin resonances correctly, the very large number of parameters
would not allow to reach any clear conclsions on the possible manifestations of 
new resonances. Notice that such resonances are expected above the first
resonance region, where higher spin resonances are expected to play significant
roles.

The advantage of the quark model for the meson photoproduction is the ability 
to introduce all known resonances. Moreover, the number of adjustable parameters,
one per resonance in the broken $SU(6)\otimes O(3)$ limit, stays much smaller
than in the case of the above formalism. Contrary to the former approache, the 
quark model adjustable parameters measure the extent to which the 
$SU(6)\otimes O(3)$ symmetry is broken. Hence, they should stay rather close
to their $SU(6)\otimes O(3)$ symmetry values, while in the case of effective
Lagrangians, apart for a few exceptions, there are no constraints on the range
of the fitted parameters.
Besides, the constituent quark models allow us to relate the data directly 
to the internal structure of the baryon resonances. 

The main conclusion here is therefore: {\it the appropriate framework in search for
new baryonic resonances is constituent quark approaches}.

The above conclusion was illustrated in this note by two examples and the
findings are:
\begin{itemize}
\vspace{-3mm}
\item{
Recent $\gamma p \to K^+ \Lambda$ SAPHIR data can be understood by
taking into account the known resonances within an
effective hadronic Lagrangian approach embodying off-shell effects for spin-3/2
baryon resonances, as well as within a constituent quark model. There is
hence no need for introducing unkown resonances.
}
\vspace{-3mm}
\item{
Investigation of the recent $\gamma p \to \eta p$ Graal data within a chiral 
constituent quark approach based on the broken $SU(6)\otimes O(3)$ 
symmetry, shows clear need for a new $S_{11}$ resonance, with mass $M \approx$ 
1.730 GeV and total width $\Gamma \approx$180 MeV.
}
\vspace{-3mm}
\end{itemize}

To gain insights to the nature of this resonance, an extension of the
$\eta$ electroproduction studies above the first resonance region
is in progress~\cite{ZSL}. Investigation of vector meson electromagnetic
production within constituent quark models~\cite{CQM-phi,Omega}
appears also very promissing in baryon spectroscopy.
\section*{acknowledgements}
%
%
%
It is a pleasure to thank the organizers for their kind invitation to this
very stimulating Memorial meeting.
I am indebted to my collaborators: C. Fayard, G.H. Lamot, Z. Li,  
T. Mizutani, P. Oswald, F. Tabakin, T. Ye, and Q. Zhao.

%
 
%
\end{document}